\documentclass[journal]{IEEEtran}

\usepackage{epsfig,makeidx,color}
\usepackage{amsmath,amssymb,bbm}
\usepackage{cite,graphicx,lipsum}

\usepackage{subcaption}
\usepackage{tabularx}
\usepackage{float}
\usepackage{booktabs}
\usepackage{amsthm}
\usepackage{enumerate}
\usepackage[switch,pagewise]{lineno}
\usepackage{hyperref}

\begin{document}

\title{Artificial Intelligence Techniques for \\Next-Generation Massive Satellite Networks}

\author{Bassel Al Homssi*, Kosta Dakic, Ke Wang, Tansu Alpcan, Ben Allen, \\ Russell Boyce, Sithamparanathan Kandeepan, Akram Al-Hourani, and Walid Saad
	\thanks{B. Al Homssi and R. Boyce are with UNSW Canberra Space, K. Dakic, K. Wang, S. Kanddepan, and A. Al-Hourani are with RMIT University; T. Alpcan is with the University of Melbourne; B. Allen is with OneWeb and the University of Surrey; W. Saad is with Virginia Tech. Corresponding author: B. Al Homssi, Email: bhomssi@ieee.org.}\\
\thanks{\copyright 2023 IEEE. Personal use of this material is permitted. Permission from IEEE must be obtained for all other uses, in any current or future	media, including reprinting/republishing this material for advertising or promotional purposes, creating new collective works, for resale or	redistribution to servers or lists, or reuse of any copyrighted component of this work in other works.}}

\markboth{IEEE Communication Magazine,~Vol.~x, No.~x, May~2022}
{Shell \MakeLowercase{\textit{et al.}}: Bare Demo of IEEEtran.cls for IEEE Journals}
\maketitle

\begin{abstract}
	Space communications, particularly massive satellite networks, re-emerged as an appealing candidate for next generation networks due to major advances in space launching,  electronics, processing power, and miniaturization. However, massive satellite networks rely on numerous underlying and intertwined processes that cannot be truly captured using conventionally used models, due to their dynamic and unique features such as orbital speed, inter-satellite links, short pass time, and satellite footprint, among others. Hence, new approaches are needed to enable the network to proactively adjust to the rapidly varying conditions associated within the link. Artificial intelligence (AI) provides a pathway to capture these processes, analyze their behavior, and model their effect on the network. This article introduces the application of AI techniques for integrated terrestrial satellite networks, particularly massive satellite network communications. It details the unique features of massive satellite networks, and the overarching challenges concomitant with their integration into the current communication infrastructure. Moreover, this article provides insights into state-of-the-art AI techniques across various layers of the communication link. This entails applying AI for forecasting the highly dynamic radio channel, spectrum sensing and classification, signal detection and demodulation, inter-satellite and satellite access network optimization, and network security. Moreover, future paradigms and the mapping of these mechanisms onto practical networks are outlined.
\end{abstract}


\IEEEpeerreviewmaketitle

\section{Introduction}
Revolutionary leaps in the modern world are predicated on obtaining, retaining, sharing, and using information as efficiently as possible. Data sharing and communication are now at the forefront of data-hungry applications across all elements of a functioning society. The financial, political, and societal implications of efficiently and quickly relaying information are difficult to overstate. Despite the rapid advancements in the information economy, our existing communication infrastructure remains heavily reliant on terrestrial communications. However, current terrestrial infrastructure is at the verge of becoming solely insufficient to achieve a truly global communication system.

In the new commercial space-age, space-based communications have advanced to providing indispensable communication functions. Satellite communication systems emerged as a great candidate to supplement existing terrestrial networks, offering a pathway to true global connectivity and bridging the coverage divide~\cite{9755278}. Hence, satellite networks are expected to provide terrestrial and aerial communications, specifically unmanned aerial vehicles (UAVs), with the necessary support for backhaul in next-generation networks~\cite{8869705}. Thus, future visions of global communication and coverage are inextricably tied to the emergence of massive low Earth orbit (LEO) constellations in the leading role of integrated terrestrial and non-terrestrial networks (TNTNs).

In the past few years, we crossed pivotal milestones in the space game through the rise of pioneering private satellite networks (e.g., SpaceX, OneWeb, Amazon, etc.), in the satellite constellation design, operation, and research sectors. Current efforts to standardize next generation satellite networks are materializing, however many fundamental challenges remain~\cite{9275613}. These inherent challenges stem from the inherent characteristics of LEO satellites; their fast orbital motion, range of coverage, and changing topology when compared to conventional static geosynchronous satellites. 

Matching the demand for autonomously adaptive mechanisms with a rapidly changing radio environment, present in satellite networks, requires the synthesis of agile methods that are quick and dynamic enough to adapt at near-instant rates. Artificial intelligence (AI) techniques can transcend reactive optimization mechanisms, offering a proactive solution that is highly adaptable and technology aware. Unlike other conventional techniques, AI-enabled next generation satellite networks are able to learn nonlinear behaviors in an autonomous manner to optimally determine the necessary system configurations. With the aid of AI, satellite networks can now be equipped with the necessary tools to proactively interact with their radio environment to provide an enhanced quality-of-service (QoS).

The main contribution of this article is a synopsis of AI techniques that enable the realization of fully autonomous and robust next generation massive LEO satellite networks. The article outlines the fundamentals of massive satellite networks and their various unique features, details state-of-the-art AI techniques for channel forecasting, spectrum sensing, signal detection, inter-satellite and access network optimization, and network security in massive satellite networks, and highlights future paradigms for AI in next generation massive satellite networks.

\begin{figure*}
	\centering
	\includegraphics[width=0.7\linewidth]{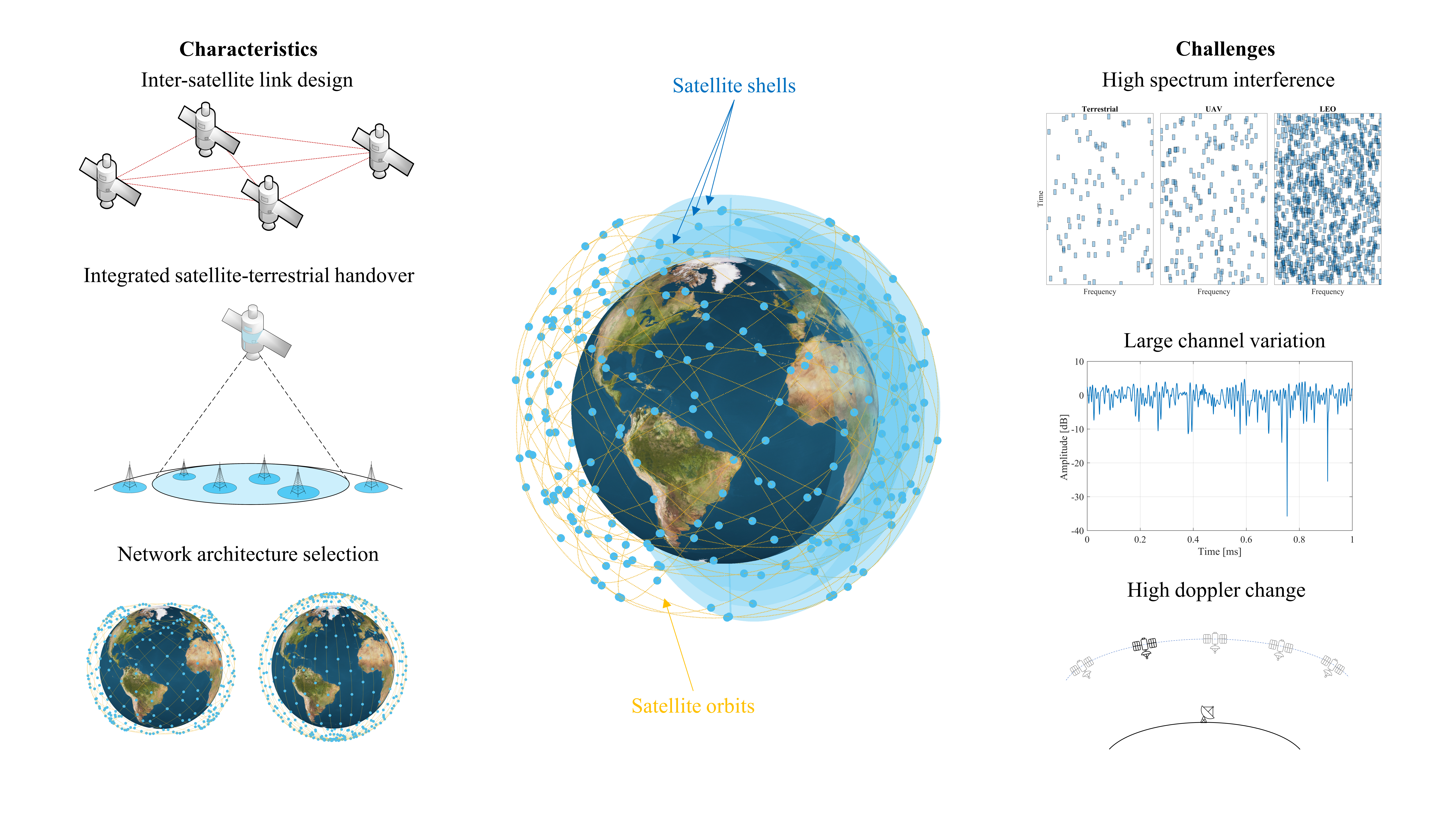}
	\captionsetup{font=footnotesize,labelfont=footnotesize}
	\caption{An overview of massive LEO satellite network and its unique features. The network consists of three different shells of satellites continuously orbiting Earth providing global coverage to its ground users.}
	\label{Fig_Overview}
\end{figure*}

\section{Features \& Challenges of Satellite Networks}
With increased demand for high throughput networking and the need for delay-sensitive applications, LEO satellites are well-positioned to support global Internet access from space~\cite{Dely_Not_Option}. Nevertheless, this comes at the expense of reduced footprint per satellite relative to GEO links. To compensate for this coverage deficit, massive satellite constellations range from a few hundreds to tens of thousands of satellites, continuously orbiting Earth, aiming to provide ground users with seamless Internet access. Hence, we categorize LEO networks' features and challenges into those incorporated with stand-alone nodes and massive constellations. Fig.~\ref{Fig_Overview} provides an overview of a typical massive LEO satellite network and details some of its unique features.

\subsection{LEO Satellite Nodes}
LEO satellites orbit Earth at very large speeds (it can take a LEO satellite 1.5 -- 2.25 hours only to complete a full orbit), making the establishment of a stable physical connection challenging, since the pass time, i.e., the time that the satellite is visible to a ground user, is extremely short (in the order of a few minutes). Furthermore, these high velocities lead to high Doppler shifts, e.g., approximately 400 kHz in the Ku-band, making signal detection and tracking challenging, especially for wideband multi-carrier systems. These high orbital velocities also lead to short channel coherent times, resulting in rapid variations in the radio channel.

Moreover, due to their high altitudes relative to aerial and terrestrial communication systems, the footprint covered by LEO satellites is typically large, thus, enabling more transmissions to appear at the receiver window. This is especially critical since satellite networks are projected to support services that rely on large numbers of users such as massive internet of things (IoT) networks. While many of those transmissions fade as the receiver altitude increases due to its sensitivity, its footprint expands which significantly increases the number of interfering transmissions it can potentially capture. Footprint and spot formation are responsible for the shaping of the radio coverage, according to user demand, which involves an intricate trade-off between coverage, interference, and complexity, while simultaneously aiming to improve spectrum utilization and link throughput.

Furthermore, satellites follow fairly predictable orbits over short periods, however, long-term they require orbital corrections and adjustments. This is because of the inherent non-uniform Earth gravitational field and solar/lunar pull, as well as the drag caused by the thin atmosphere and the solar wind. The slow drift in satellite orbits, if not corrected, will impact the network topology and performance. As a result, edge and onboard AI techniques are required to mitigate these various and often nonlinear challenges.

\subsection{LEO Satellite Constellations}
Massive LEO satellite networks are envisioned to be connected with one another using inter-satellite links (ISLs) to enable data routing and long-distance communications. Note that the vast majority of today's satellites still act as a simple bent-pipe architecture that relies on dense ground station networks. In contrast, ISLs take advantage of the free-space medium and speed of light, contrary to terrestrial links that rely on optical fiber that provide propagation speeds up to two thirds the speed of light. Therefore, massive LEO satellite networks are great candidates for not only replacing GEO links but also providing a lower latency alternative to terrestrial fiber-based Internet~\cite{Dely_Not_Option}. Nevertheless, ISL communications require accurate satellite navigation and routing optimization techniques which are often challenging to attain.

\begin{figure*}
	\centering
	\includegraphics[width=0.45\linewidth]{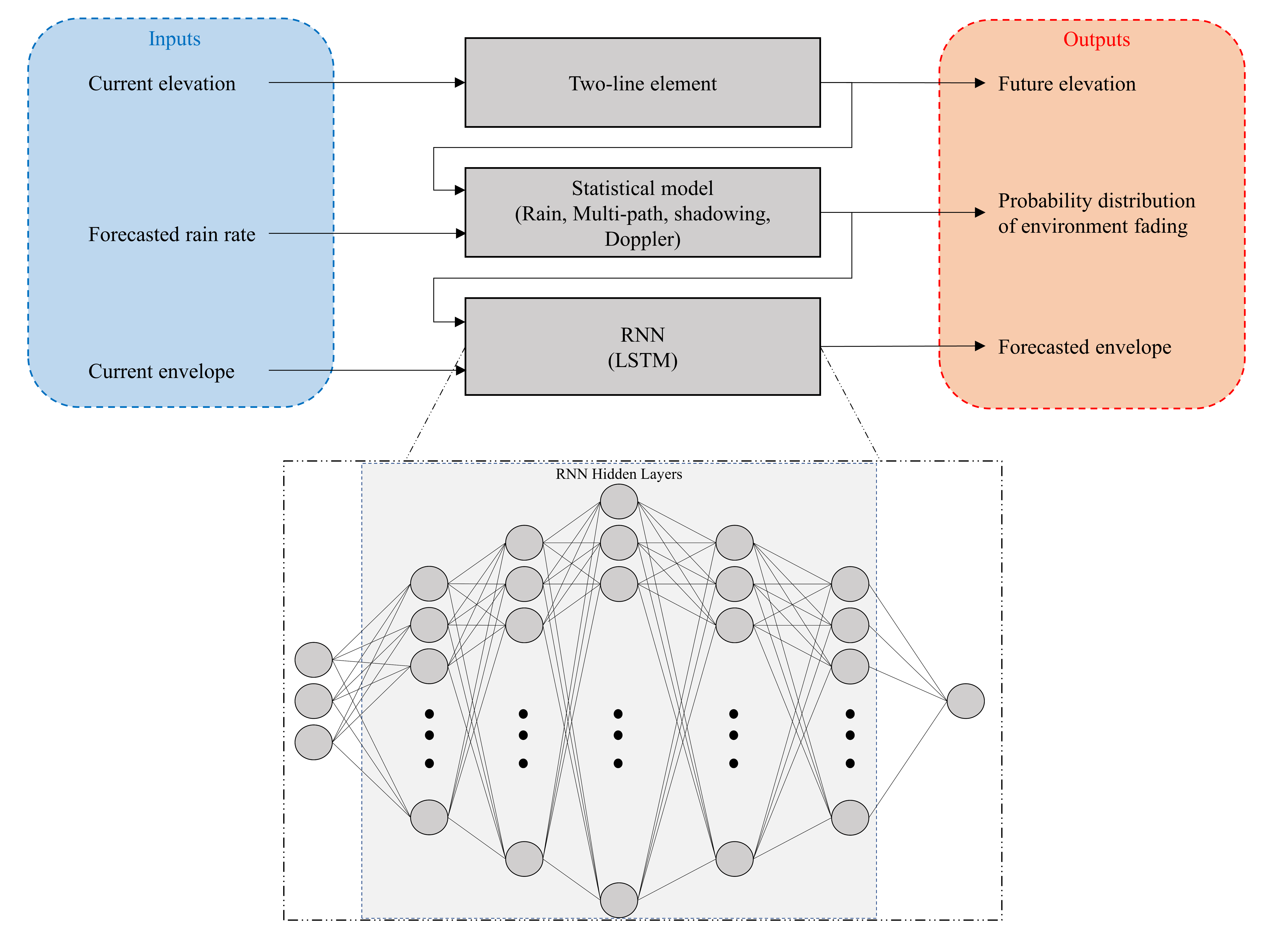}
	\captionsetup{font=footnotesize,labelfont=footnotesize}
	\caption{An example of a channel forecasting hybrid model architecture that integrates the two-line element, statistical model, and recurrent neural network. The combination provides more accurate prediction for short and long-term forecasting~\cite{10153617}.}
	\label{fig_ChanPred}
\end{figure*}

An inherent advantage of deploying massive number of satellites in the constellation is the enhancement in the network redundancy and capacity. However, as the number of satellites in the constellation increases, challenges such as network management, access network optimization, and network security become more glaring. Another key challenge is the standardization and harmonization between terrestrial networks, such as 5G/6G, and non-terrestrial networks, which exacerbates the complexity in achieving seamless integrated TNTNs~\cite{9755278}.

Massive satellite networks exhibit underlying trends that are challenging to capture using common methodologies, especially as the network is growing in size with the constant addition of more satellites and ground terminals as the network is scaled. To address these constantly varying challenges, AI techniques that rely on real-time learning can provide dynamic and proactive methodologies to achieving harmony and optimization in TNTNs. The major advantages of AI techniques are their inherent ability to be retrained to adapt to scalability and growth, as well as, the ability to transfer enabling the sharing of unique information among the satellites in the constellation using distributed learning techniques. This enables the network to be more robust and adaptive to variations that may occur as the operating environment is changing. In Sec.~\ref{Sec_Network}, we detail the deployment of several state-of-the-art AI techniques to address ISL optimization, satellite access network, and network security.

\section{On-Board AI for Satellite Communications}\label{Sec_Onboard}
\subsection{Channel Forecasting}
The major proportion of satellite signal propagation is under free-space conditions and only experiences elevated levels of fading due to two main factors: (i) interaction with the \textit{clutter} near ground causing multi-path attenuation, and shadowing, and (ii) \textit{atmospheric absorption} and \textit{rain fading}. Generally, these interactions impose significant variations on the signal envelope and deteriorate the link's QoS. Moreover, next-generation satellite links are expected to operate in higher millimeter wave frequency bands (e.g., Q/V, and even W-band) to enhance link capacity and tackle the currently heavily crowded C/Ku/Ka-bands. However, the signal at these high frequency bands suffers not only from increased fading but also from increased variability, due to rainfall and variability in air density, which is difficult to predict using conventional methods (e.g., ITU P.618). Hence, it is crucial for the medium access control (MAC) layer to proactively predict and accommodate these variations by employing different strategies such as proactive power control, frequency selection, modulation coding scheme control, and footprint/gain formation.

Due to the rapid relative movement of LEO satellites, the elevation angles seen at a particular ground station are continuously varying. Inherently, the signal envelope variations are also impacted by the elevation angle, whereby both the clutter and the atmospheric effects are interacting differently at different angles. While the elevation angle is predictable on the short-run and can be approximated by estimating the pass of the satellite, using its two-line element (TLE), the channel variations (due to elevation angle change) are challenging to predict. These variations are commonly captured using long-term statistical and regression models that are limited in providing the accurate time-series prediction required for the MAC operation. This is because these models excel in long-term prediction, which is less relevant for LEOs, since their pass is typically shorter than 20 minutes. Current practical deployments of LEO links are reactive and do not implement any forecasting techniques, thus the link would completely fail in cases that involve rainfall for instance, before adjusting.

Using AI, the network can potentially, in real-time, forecast and adjust its parameters to provide seamless service. Recurrent neural networks (RNN) are a potential candidate for time-series prediction due to their inherent sequence detection and prediction capabilities. Recently, long short-term memory (LSTM), was deployed to forecast the effect of rain fading in the Q/V-band, and was shown to achieve relatively accurate short-term prediction, but starts to deviate as the prediction window enlarges~\cite{10153617}. Nevertheless, the accuracy of medium-term prediction can be significantly enhanced by incorporating empirical-based statistical modeling in the training of the neural network, as illustrated in Fig.~\ref{fig_ChanPred}. For the purpose of distributed and onboard learning, new classes of neural networks are emerging that promotes parallelization and reduces the complexity of the distributed networks. For example, AI transformers are a novel concept that entails a self-attention property which elevates certain data features while diminishing others making it more suitable for distributed learning.

Although LSTMs and transformers provide suitable prediction, they require large amounts of measurements in order to adequately capture the effect of the elevation angle on the underlying processes. For instance, if the rainfall profile varies significantly, then these techniques will not be able to rapidly adapt due to the short rain burst duration. Moreover, for applications that involve intermittent transmissions, such as massive IoT sensors, obtaining such continuous measurements can be energy inefficient. Therefore, potential hybrid techniques that rely on smaller portions of training measurements while maintaining the accuracy are needed to facilitate efficient and continuous channel forecasting.

\subsection{Spectrum Sensing and Classification}
\begin{figure}
	\centering
	\includegraphics[width=0.75\linewidth]{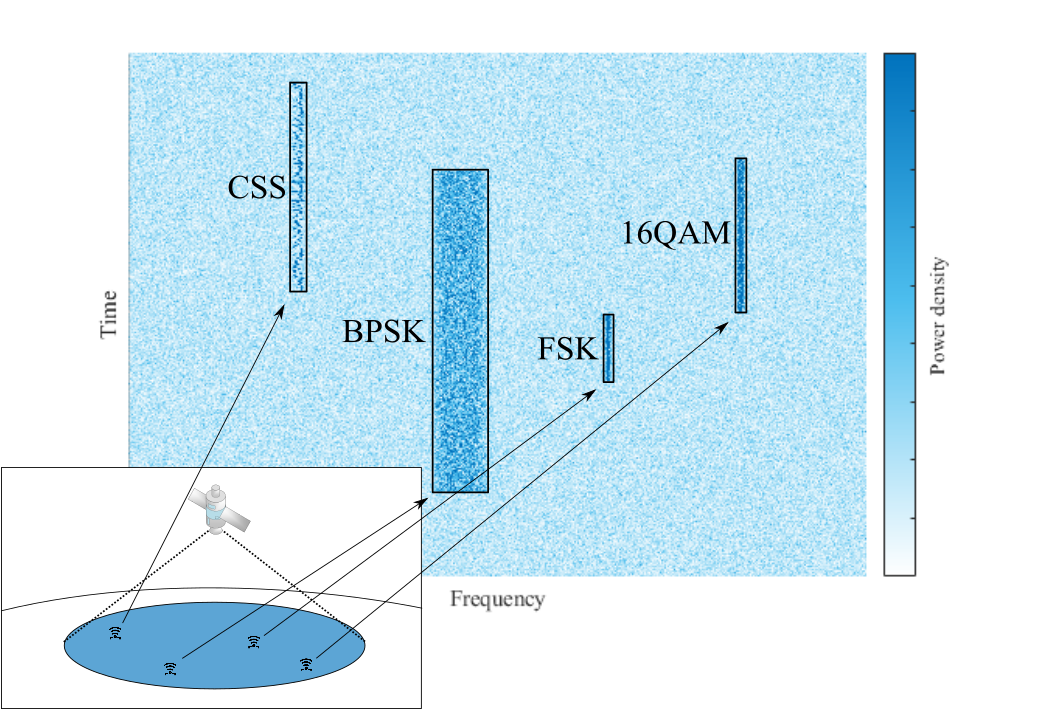}
	\captionsetup{font=footnotesize,labelfont=footnotesize}
	\caption{Spectrum classification for non-homogeneous traffic. Traffic shown composes of chirp spread spectrum (CSS), binary phase shift keying (BPSK), frequency shift keying (FSK), and quadrature amplitude modulation (QAM).}
	\label{fig_Spectrum}
\end{figure}

Due to the increasing number of \textit{broadband} satellite constellation providers, radio spectrum bands are reused and shared among these different operators. These operators are anticipated to have different technologies with assorted data rates, bandwidths, and frame duration, as illustrated in Fig.~\ref{fig_Spectrum}. Understanding the spatio-temporal availability of radio resources is particularly important to combat cross-constellation interference. Transmitting in a congested spectrum band would potentially deteriorate signal reception due to high levels of interference, resulting in signal collisions and packet losses. Thus, spectrum sensing and frame classification aids in optimizing radio spectrum allocation in both the uplink and downlink, thus enhancing the communication success rate in cross-constellation shared radio bands. 

In addition, \textit{narrowband} LEO satellite constellations typically use the lower frequencies in the UHF and S-band to carry IoT sensor traffic, this is due to the favourable non-line-of-sight tolerance of these bands, and due to the ability to use omni-directional antennas at the ground terminals. These bands are typically very busy and prone to interference due to the large number of connected devices and can benefit from a good spatio-temporal modeling of the radio spectrum. Modeling is critical when deploying in the already busy license-free spectrum bands that can be typically shared with terrestrial IoT applications~\cite{9755278}.

A classical method of spectrum sensing involves operating a spectrum analyzer to sweep over the frequencies followed by energy detection to obtain the occupancy. However, the limited temporal resolution of the spectrum analyzer and bursty nature of massive IoT traffic make such frames difficult to detect and identify, which could result in a heightened probability of misdetection and false alarm. Other modeling techniques that are heavily used in spectrum modeling rely on first-order hidden Markov models to define and reduce the detection error probabilities. Nevertheless, hardware limitations, nonuniform traffic, and different spectrum bands feature different sets of challenges which make it quite challenging to accurately represent the spectrum occupancy using these first-order models.

Spectrum modeling and frame identification rely on a two-step machine learning (ML) technique; firstly, spectrum occupancy sensing is employed to detect frames and mark them with bounding boxes in time-frequency and secondly, automatic modulation classification is envisioned with a classification network such as a convolutional neural network (CNN). Spectrum sensing and classification could be further enhanced by using various AI techniques, which are less dependent on expert feature selection. Region-based CNN (RCNN) or YOLO hold potential to be used in detecting the beginning, duration, bandwidth and center frequency of each frame, along with the ability to accurately classify the modulation type and technology~\cite{9075413}.

\subsection{Signal Detection}
\begin{figure}
	\centering
	\includegraphics[width=0.75\linewidth]{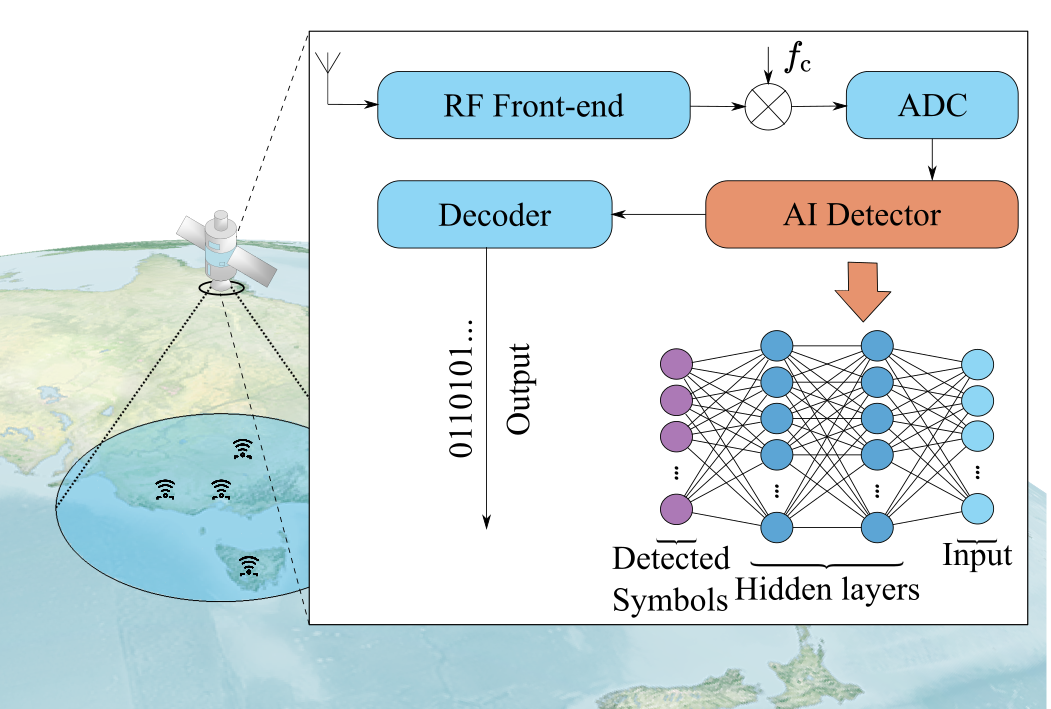}
	\captionsetup{font=footnotesize,labelfont=footnotesize}
	\caption{AI signal detection and demodulation for a satellite link.}
	\label{fig_SigDetect}
\end{figure}

Efficient signal detection is a requirement for communications, particularly in the context of massive satellite networks with a vast number of satellite-to-ground connections. It has been shown that using AI for signal detection in communication systems can outperform conventional approaches~\cite{10107653} under interference. Conventional signal detectors typically rely on tractable mathematical models under a known noise process and/or deterministic interference. On the other hand, AI approaches can be effectively used under dynamic interference to effectively detect the target signal. In addition, an AI detector can be trained for detecting various modulation and coding techniques. However, a higher-order modulation scheme will incur an increased training data set size due to the large symbol set size (for instance, 1024-QAM has 1024 different symbols), which might limit the practicality of using conventional training methods and exponentially increase the neural network complexity. An illustration of an AI-based signal detector is presented in Fig.~\ref{fig_SigDetect}.

Different AI architectures such as multi-layer perceptron (MLP), CNN, and RNN can be employed for signal detection~\cite{OShea}. MLPs are substantially less powerful than the other two architectures, hence the majority of research focuses on CNN and RNN for signal detection. CNN requires a fixed-length input, thus a symbol-by-symbol detector framework can be realized. However, since CNN can only perform symbol-by-symbol detection, adjacent symbols within the signal are not accounted for. A sequence detector network using an RNN network can be employed to combat the limitation of the symbol-by-symbol detector. The complexity of those detectors is a challenge, particularly when the detector is deployed on a satellite and energy efficiency is a priority.

Spiking neural networks (SNNs) can be utilized to tackle this issue, as they are much more energy-efficient relative to traditional artificial neural networks (ANNs) and therefore reduce the power consumption requirement which is especially crucial in satellites which rely on limited solar power. Contrary to traditional ANNs which use a constant value to activate neurons, SNNs take inspiration from a biological system and activate neurons with a series of spikes~\cite{RN660}. By adding a spatio-temporal dimension, SNNs are effective at processing continuous data streams, such as wireless signals.


\section{Network AI for Satellite Communications}\label{Sec_Network}
\subsection{ISL Communications Optimization}
ISLs provide the interconnection between satellites to achieve data transfer and ranging in space. ISLs allow the cooperation among satellites, and hence, reduce the stringent limitations on the terminals. In ISL connected networks, end-to-end satellite-based traffic routing enhances throughput and reduces the communication latency. Conventional microwave, millimeter-wave, terahertz-wave, and optical wave have all been proposed to enable ISLs, with the objective of achieving high data rate, stable links, and long interconnection distance.

Whilst ISLs provide significant benefits to satellite networks, due to the high moving speed of LEO satellites, the visibility of neighboring satellites is limited and rapidly changing, resulting in a time-varying satellite network topology. Also, the constraints on the number of ISL ports, operation band and beam patterns, further limit the interconnection availability and capability. Hence, optimal scheduling and resource allocation of ISLs among satellites is required. However, it is highly challenging to solve these limitations, where prediction-based scheduling and resource allocation methods using historical data for proactive management are highly promising.

Existing prediction-based studies (e.g., \cite{ISL_ML_review} and references therein) have mostly focused on optimizing the use of ISL resources to maximize the capacity and/or minimizing the date interconnection density. While these studies have achieved significant improvement in ISL performance, the solutions are not designed for highly dynamic traffic patterns, such as IoT-over-satellite packets, where satellite is expected to support crowded regions. To solve this problem, traffic prediction-based methods has also been further proposed to learn the traffic pattern and divide the interconnection resources more efficiently, leading to task-driven methods~\cite{ISL_small_traffic}.

Reinforcement learning (RL) techniques have also been widely studied in ISLs to optimize the traffic scheduling and resource allocation for satellite-to-satellite communications. The deep Q-network is a popular option, which is a value-based method that matches the utility function and actions under different states. However, since the action space in ISLs is high dimensional and continuous, the objection function value computation cost is normally high. Hence, the policy-based RL method is also widely explored, where the policy gradient for actions is calculated for optimization. Combining the value-based and the policy-based algorithms, the \textit{actor-critic framework} has been applied in ISLs~\cite{ISL_RL2}. However, distributed RL has the fundamental limitation of network convergence, while centralized RL relies on high computation power resulting in high ISL communication resource consumption.
By using ISLs, traffic offloading, coverage extension, capacity enhancement and latency reduction have been realized. Another promising solution is the use of \textit{experienced RL agent} method \cite{ISL_latency}, where a generative adversarial network (GAN)-based refiner allows the RL agent to gain network experience in a virtual environment before real deployment to achieve both high reliability and low latency.

Whilst ML algorithms can solve complex multi-domain optimization problems, the added complexity to ML models and computations needs to be considered. This is particularly when incorporating the different satellite communication properties in these bands and other physical limitations. For example, optical ISL can provide very high-speed connections between satellites, however, it is highly vulnerable to minor misalignment caused by orbital perturbations. This requires the ML algorithm to be equipped with rapid adaptation and update. In addition, ISLs are an integral part of a satellite constellation, and hence, the scheduling and resource allocation need to be co-designed and co-optimized with other access traffic.

\subsection{Satellite Access Network Optimization}
The planning, deployment, and operation of wireless networks require making continuous trade-off decisions to optimize resource allocation and maximize performance. Classical problems in this broad domain include routing, radio resource allocation, interference management, wireless power control, and terminal node energy optimization, among others.  In the context of massive satellite constellations, resource allocation decisions are of a great importance due to the complex network topology adding to the continuous variations in satellite and ground stations available. This dynamic availability nature of satellite network nodes results from many factors including the inherent occlusion due to Earth curvature, the limitation of satellite footprint, and the limitation in available ISL ports. Convex and distributed optimization methods as well as game-theoretic approaches have been used by the research community over the last few decades to address similar multi-faceted resource allocation challenges in wireless networks~\cite{alpcan_boche_honig_poor_2013}. These methods are today part of the toolbox of modern wireless and satellite network designers. 

\begin{figure}
	\centering
	\includegraphics[width=\linewidth]{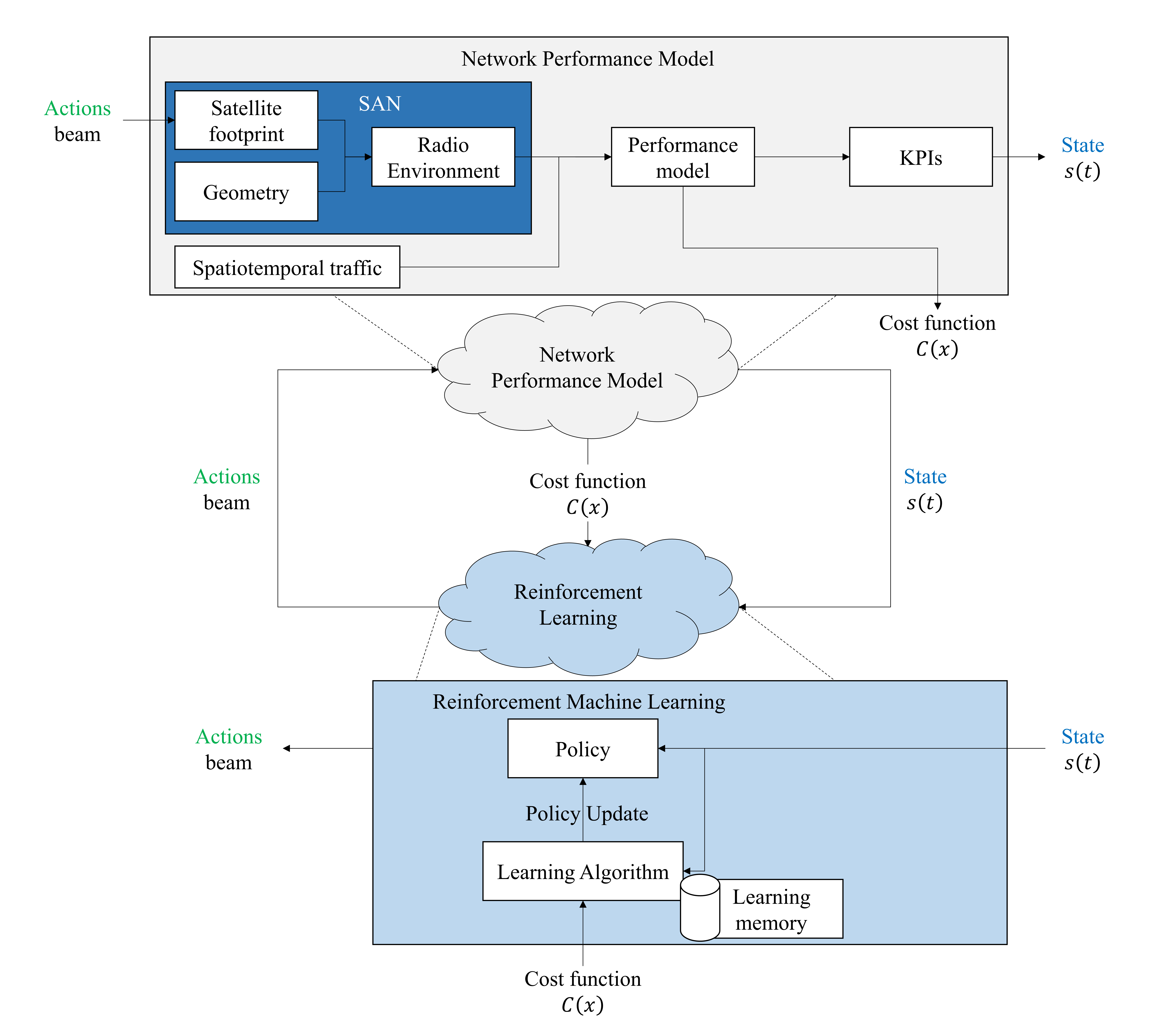}
	\captionsetup{font=footnotesize,labelfont=footnotesize}
	\caption{AI for SAN optimization in massive satellite networks.}
	\label{fig_ITSN}
\end{figure}

However, classical optimization methods require detailed models, which often require extensive manual engineering efforts. In contrast, data-oriented AI approaches can provide the much needed flexibility and automation to addressing resource allocation problems in dynamic wireless and satellite communication networks. Here, once again, RL can be utilized instead of classical optimization models to learn complex and dynamic relationships between network parameters and the target QoS and efficiency levels. RL techniques can be exploited in satellite access networks (SANs) as illustrated in Fig.~\ref{fig_ITSN}. The SAN model would include inputs such as the beamwidth in the form of footprint shaping. A contributing factor is the spatiotemporal spectral traffic. On the other hand, the outputs of the model are fed to the RL model as the cost function in the form of the required key performance indicators (KPIs) that determine the state of the SAN, such as the statistics of data throughput or the SNR. RL techniques are a good fit for highly dynamic radio channels that are impacted by varying conditions such as in SANs.

In addition, dynamic optimization of network routing strategy can significantly improve latency performance. Traditionally, the shortest-path algorithm is used for routing. However, the shortest link is not always the most efficient, or stable in such dynamic system. 
Combining AI with optimization in SANs poses unique challenges. Firstly, solutions of data-oriented methods heavily depend on the quality and quantity of network management data. Therefore, the quality of the data collected from the SAN or its digital twins should be carefully checked, also a careful balance in performance metrics reporting, i.e. data quantity, is needed to avoid increased overhead. Moreover, most contemporary ML algorithms and optimization methods focus on performance and do not sufficiently emphasise robustness. Paradoxically, in many practical cases, robustness matters more than accuracy or efficiency.

\subsection{Network Security}
The inherent broadcast nature of wireless networks make them more vulnerable to jamming, data theft, and spoofing when compared to wired networks. This is only exacerbated with the wide footprint of satellite links. Network security is typically addressed separately in each of the TCP/IP layers; from the network access (PHY/MAC) layer, network layer, transport layer, to the application layer. A great lesson from past communication technologies developed over the decades is to have open and harmonized security protocols and encryption methods, this helps early detecting and fixing any weaknesses by the large development community, however, current satellite communication networks seem to heavily rely on proprietary communication techniques that can increase the difficulty of integration with terrestrial networks. 

As the security of the upper network layers in satellite can be adopted from terrestrial networks, more innovation is required to secure the physical layer of massive satellite networks. One such methods is the use of AI to intelligently introduce narrow nulls in the coverage footprint to suppress ground-based jammers. Federated learning can be deployed to enable information sharing between satellites, within the constellation, about the spectrum, e.g., detection of certain jammers/threats at certain geographical locations. This can also be extended through the use of swarm learning which does not require a central server/node. There are also emerging approaches to exploit transmitter-specific RF fingerprints to validate the authenticity of devices based on deep learning methods~\cite{Da_paper}. Hence. even if the contents of a radio frame are not altered, the nuance changes in the I/Q vectors due to hardware imperfection leaving a detectable trace that allows distinguishing authentic from fraud (or spoofing) devices.

Using the large broadcast footprint to the satellite advantage, quantum key distribution (QKD) can be effectively employed in dense satellite networks. Using properties from quantum mechanics, the parties sharing quantum keys can detect any eavesdropper trying to access the transmission. Using One-time pad algorithm with QKD can produce provably secure (unbreakable) communication between the satellites. The coupling of quantum coding with AI techniques, quantum AI, results in a more versatile system that enables the defender to more accurately and rapidly detect any novel security threats. This is achieved as the network becomes smarter as more data is being learned in real-time. Moreover, quantum AI is characteristic of high learning rates, exploited through quantum neural networks which utilize quantum processing. However, quantum AI is still in its infancy stages and requires more research iterations to be utilized in practical networks.

\section{Conclusion}
Current traditional network optimization techniques cannot achieve efficient, autonomous, and harmonious integration. Hence, we provided an overview of various AI-enabled techniques for satellite networks to achieve enhanced resource utilization and robust services. The presented techniques deal with the satellite network across its various components targeting characterization, forecasting, and optimization for both edge nodes and the network. While current AI techniques are very promising and are transferable across technologies, many challenges emerge such as the increased complexity of the system, the reliance on large datasets and training duration. Moreover, energy constraints, due to the the reliance on solar as the sole source of energy and the lack of power-efficient and high performance processors to execute AI algorithms, restrict the usage of AI techniques in-orbit. Hence, new AI algorithms and specialized hardware are necessary to achieve resilient AI-enabled satellite constellations.

\ifCLASSOPTIONcaptionsoff

\fi
\bibliographystyle{IEEEtran}
\bibliography{Satellite_ML_Magazine.bib}

\begin{IEEEbiographynophoto}{Bassel Al Homssi}
	is currently a Lecturer at UNSW, Australia. His research interests are on wireless communications, satellite networks, and applying AI in space systems.
\end{IEEEbiographynophoto}
\begin{IEEEbiographynophoto}{Kosta Dakic} 
	is a Ph.D. candidate at RMIT University, Australia. His research interests include wireless communications, IoT, and the application of deep learning in telecommunications.
\end{IEEEbiographynophoto}
\begin{IEEEbiographynophoto}{Ke Wang}
	is an Associate Professor at RMIT University. His research interests mainly include optical and wireless communications and convergence, machine learning in telecommunications, and integrated opto-electrronic devices and circuits. 
\end{IEEEbiographynophoto}
\begin{IEEEbiographynophoto}{Tansu Alpcan}
	is currently with the Dept. of Electrical and Electronic Engineering at The University of Melbourne as a Professor and Reader. His research interests include applications of control, optimization, and game theories, and machine learning to security and resource allocation problems in communications, smart grid, and IoT.
\end{IEEEbiographynophoto}
\begin{IEEEbiographynophoto}{Sithamparanathan Kandeepan}
	is a Professor in Telecommunications Engineering, RMIT University. His research interests are in wireless and satellite communications, he currently leads the satellite communications program for the Space Industry Hub, works with Smart Satellite Cooperative Research Centre and leads the Wireless Innovation Lab (WiLAB).
\end{IEEEbiographynophoto}
\begin{IEEEbiographynophoto}{Ben Allen}
	is Director of Communication Systems Engineering at OneWeb and visiting Professor with the University of Surrey, UK. He has been researching radio communications for several decades, spanning both academia and industry.
\end{IEEEbiographynophoto}
\begin{IEEEbiographynophoto}{Russell Boyce}
	is currently the Director of UNSW Canberra Space. His research interests are in hypersonics, AI-enabled space systems, and satellite networks.
\end{IEEEbiographynophoto}
\begin{IEEEbiographynophoto}{Akram Al-Hourani}
	is an Associate Processor at the School of Engineering, RMIT University. He has extensive industry/government engagement as a chief investigator in multiple research projects related to Satellite Communications, The Internet-of-Things (IoT), and Smart Cities.
\end{IEEEbiographynophoto}
\begin{IEEEbiographynophoto}{Walid Saad}
	is a Professor of Electrical and Computer Engineering at Virginia Tech. His  research interests include wireless networks, machine learning, game theory, cybersecurity, unmanned aerial vehicles, cellular networks, and cyber-physical systems. He was the author/co-author of eleven conference best paper awards  and of the 2015 and 2022 IEEE ComSoc Fred W. Ellersick Prize. He is a Fellow of the IEEE.
\end{IEEEbiographynophoto}

\end{document}